\documentclass{PoS}
\usepackage[utf8]{inputenc}
\usepackage{amsmath,amssymb,amsfonts}
\usepackage{macros}
\usepackage{graphics}
\RequirePackage{bm}
\RequirePackage{bbm}

\graphicspath{{./plots//}}
\newcommand\liverpool{Department of Mathematical Sciences, University
  of Liverpool, Liverpool L69 7ZL, UK  \\
Centre for Mathematical Sciences, Plymouth University, Plymouth PL4 8
AA, UK. 
}

\title{Tackling the sign problem with a moment expansion and
  application to Heavy dense QCD 
\thanks{This work is supported by the Leverhulme Trust, grant
  RPG-2014-118 and by STFC grant ST/L000350/1.} }
      
\ShortTitle{Moment expansion and application to Heavy dense QCD}

\author{\speaker{Nicolas Garron} and 
  {Kurt Langfeld}
  \\
  \liverpool\\
  E-mail: \email{ngarron@gmail.com}, \email{Kurt.Langfeld@liverpool.ac.uk}
 }

\abstract{
Heavy-Dense QCD (HDQCD) is a popular theory to investigate the sign problem
in quantum field theory. Besides its physical applications,
HDQCD is relatively easy to implement numerically: the fermionic degrees of freedom
are integrated out, and the fermion determinant factorises into local
ones. The theory  has a sign problem, the severeness of which depends
on the value of the chemical potential, which makes this theory ideal
to test the reach of new algorithms. 
We use the LLR approach to obtain the probability distribution of the
phase of the fermion determinant. Our goal is the calculation of the
phase factor expectation value, which appears as Fourier transform of
this probability distribution. We here propose a new and systematic
moment expansion for this phase factor. We compare the answer from the
moment expansion order by order with the ``exact'' answer.  We find
that this expansion converge quickly and works very well in the strong sign problem
region. 
}

\FullConference{34th annual International Symposium on Lattice Field Theory\\
		24-30 July 2016\\
		University of Southampton, UK}

\begin{document}

\section{Heavy-Dense QCD}
This part is based on our recent work~\cite{Garron:2016noc}.
Recent progress using a complex Langevin approach have
been reported in this conference~\cite{Aarts:2016nju},
based on~\cite{Aarts:2014fsa,Aarts:2015yba}.
See also~\cite{Rindlisbacher:2015pea} for a study based on a mean-field approximation.
We start with the partition function of QCD
written as an integral over $SU(3)$ matrices 
(we choose the  Wilson action for $S_\mathrm{YM}[U]$):
\be
Z(\mu) = \int {\cal D }U_\mu  \; \exp \{ \beta \, S_\mathrm{YM}[U] \}\; 
\hbox{Det}\, M(\mu) \;.
\label{eq:ZQCD}
\ee
HDQCD is expected to be a good approximation (at least qualitatively)
of QCD if $\mu,m\gg T$. Here, $\mu$ is the chemical potential, $m$ the
quark mass and the temperature $T$ is given by the inverse of number
of lattice sites in the temporal direction, $aT=1/N_t$.
In this regime, the fermionic determinant can be approximated by 
\be
\hbox{Det} \, M(\mu) \simeq \prod _{\vec{x}} \; {\det }^2 \Bigl( 1 \,
+ \, \, \e ^{(\mu -m) / T} \, P(\vec{x}) \Bigr) 
\;,
\label{eq:deMapprox}
\ee
$P(\vec{x})$ is the Polyakov line starting at position $\vec{x}$
and winding around the torus in temporal direction: 
In general this determinant is complex, but there are three special cases
for which the determinant is real, either exactly or to a very good approximation,
namely if $\mu=0,m,\infty$.
From Eq.~\ref{eq:deMapprox}, we also observe that
$ \hbox{Det} \, M(\mu+m)  \simeq \e ^{(\mu +m) / T}  M(m-\mu) $ .
Therefore we only study the theory for $0 < \mu \le m$ and we identify three regimes
\begin{enumerate}
\item Small $\mu$, where the theory is almost real,
  the sign problem is weak.
\item Intermediate $\mu$, the theory exhibits a strong sign problem.
\item Dense regime: $\mu$ close to $m$, where again the sign problem becomes weak.
\end{enumerate}
We performed a simulation on a $8^4$ lattice with $\beta=5.8$ and mass
$am\simeq 1.427$.
With these values of the parameters, we find that the first two regions are rather
broad: for $\mu < 1.1$ the theory is real to good approximation,
and re-weighting techniques give reliable results.
The strong sign problem regime is somewhere between $1.1$ and $1.4$;
although the transition between the first two regions is rather
smooth, after $\mu\sim 1.3$ the phase of the determinant changes
rather rapidly and reaches the dense regime,
which is a rather narrow region, roughly between $1.4$ and $1.427$.

\subsection{Density of states and reweighting}
Denoting by $\phi$ the phase of the determinant
${Det} M[U] = |{Det} M[U] | \;\exp\{i\phi[U] \}$,
we have
\be
Z = \int ds \; \rho(s) \; \exp \{ i  s\}   \;
  = \int ds \; \rho(s) \; \cos(s)  \;,
\ee
 where, generalising the density of states, we introduced
  \be
\rho(s) = \int {\cal D }U_\mu  \; \exp \{ \beta \,S_\mathrm{YM}[U] \}
|\hbox{Det} M[U]| \delta(s - \phi[U]) \;, 
\ee
where  $\phi[U] $ is the sum of the phases of all the local
determinants. We also introduce the {\em phase quenched}
theory by 
\bea
\label{eq:Zpq}
Z_{PQ} & = & \int {\cal D }U_\mu  \; \exp \{ \beta \,
S_\mathrm{YM}[U] \}\; \vert \hbox{Det} M[U] \vert  
=
\; \int ds \; \rho(s) \;. 
\eea
The expectation value of an observable $X$ in the full and in the
phase quenched theory is given by
\bea
\la X \ra     &=& \frac{1}{Z} \int {\cal D }U_\mu  \; X
\exp \{ \beta \, S_\mathrm{YM}[U] \} \hbox{Det} M[U] \;, \\
\la X \ra_{PQ} &=& \frac{1}{Z_{PQ}} \int {\cal D }U_\mu  \; X
\exp \{ \beta \, S_\mathrm{YM}[U] \} |\hbox{Det} M[U]| \;, \nn\\
\eea
so that
$
\la X \ra =  \frac{ \la X \e^{i \phi}\ra_{PQ}}{ \la \e^{i \phi}\ra_{PQ}}
$
and in particular
$
Z \; = \; Z_{PQ} \; \la \e^{i\phi} \ra_{PQ} \;.
$
This reweighting approach requires the determination of the phase factor expectation value,
which is the central quantity in this work:
\be
\label{eq:phase1}
\la \e^{i\phi} \ra_{PQ}  =
\frac{\int ds \; \rho(s) \cos(s)}{\int ds \; \rho(s)}\;,
\ee
and relates the ``full'' quark density to the one defined in the
phase quenched approximation
\be
\sigma = \frac{T}{V} \frac{\partial \ln Z }{\partial \mu}  
= \frac{T}{V} \left\{
  \frac{\partial \ln \la \e^{i\phi} \ra_{PQ} }{\partial \mu}
  +  \frac{\partial \ln Z_{PQ} }{\partial \mu}   
  \right\}\;.
\ee

\subsection{LLR}

\begin{figure*}[t]
  \begin{center}
  \begin{tabular}{cc}
  \includegraphics[height=6cm]{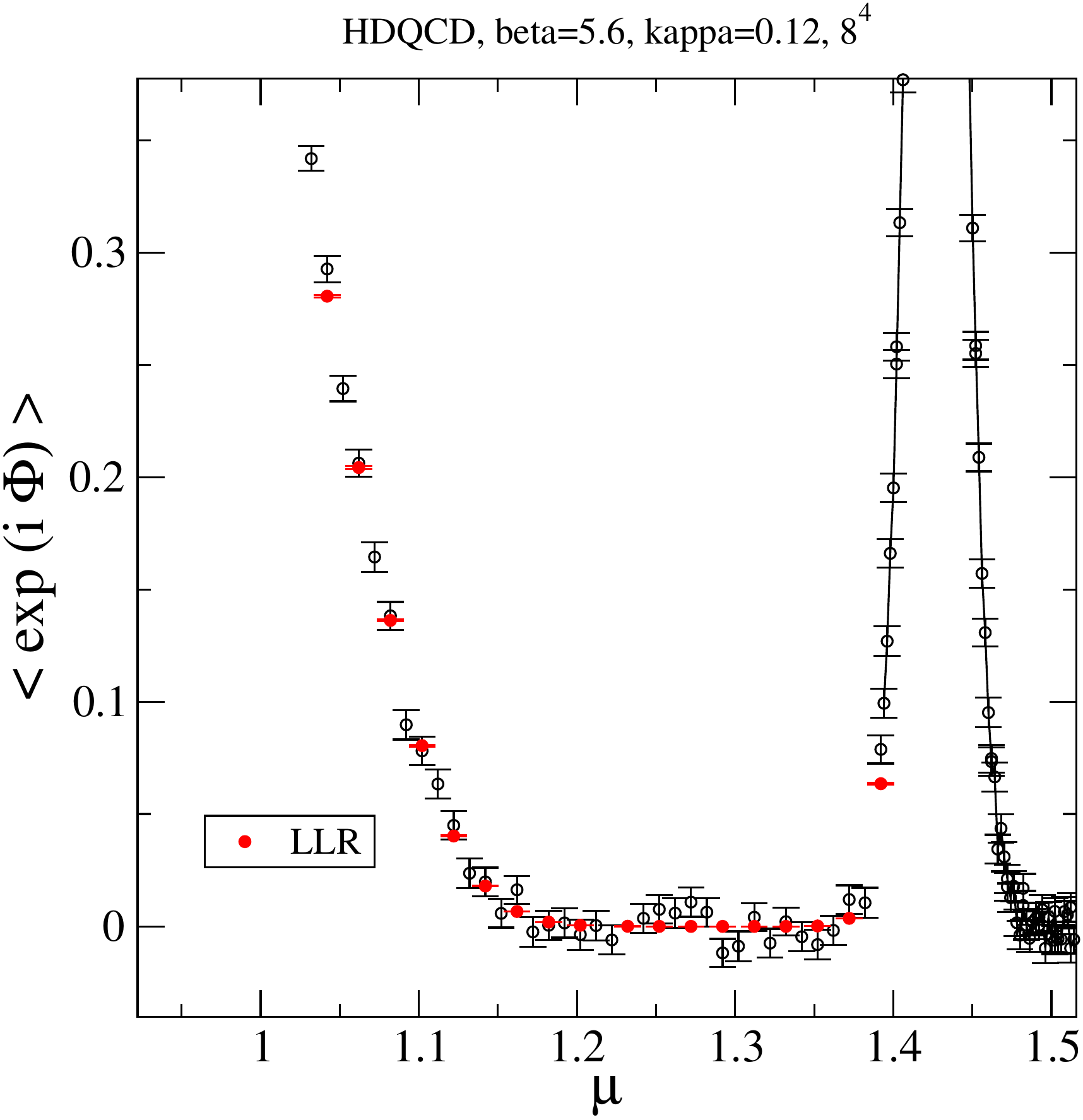} \hspace{0.5cm}
  \includegraphics[height=6cm]{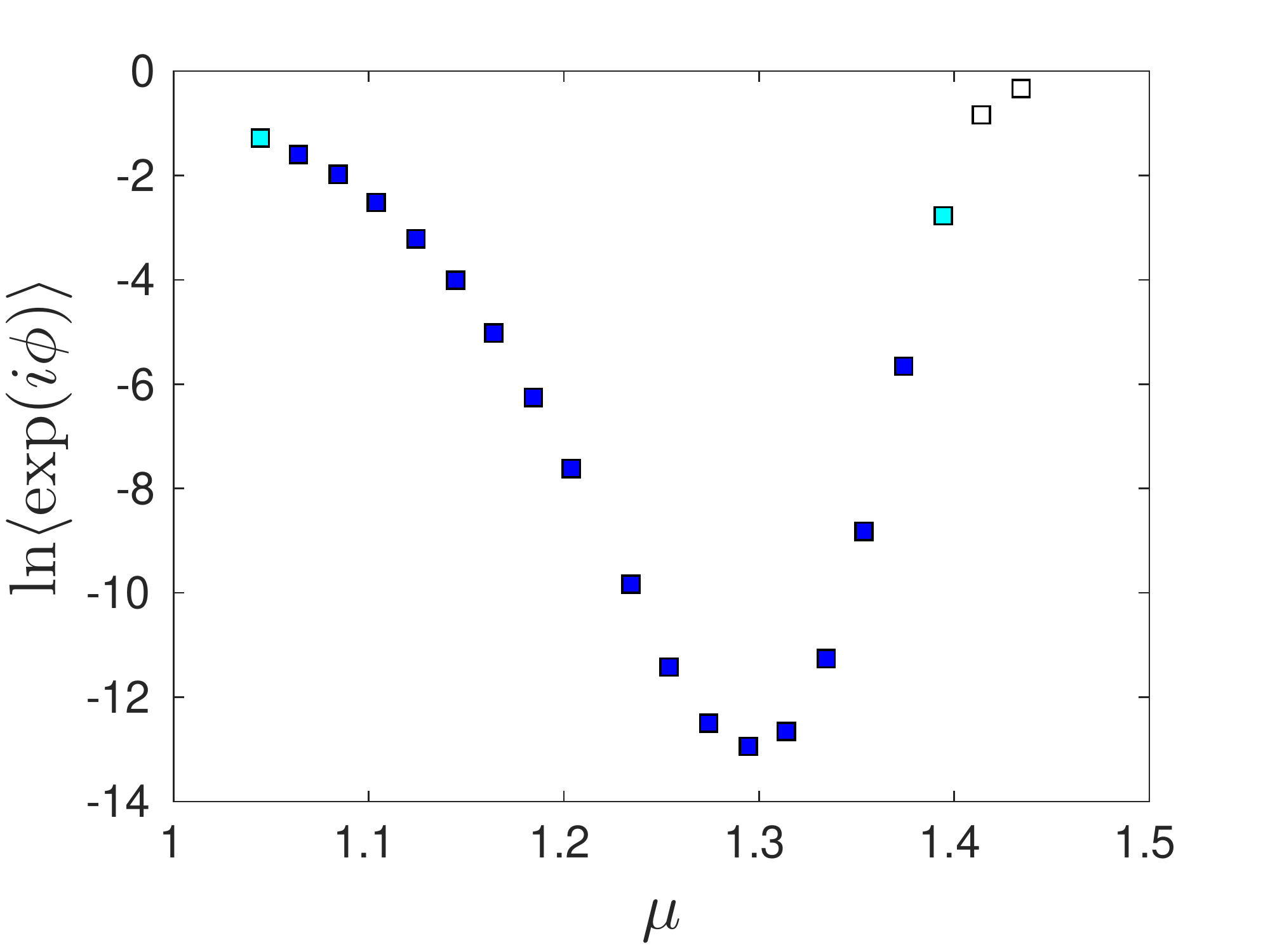}
  \end{tabular}
  \end{center}
\caption{\label{fig:phase} Left: The phase factor expectation value
  $\langle  \e^{i\phi} \rangle $ as a function of the chemical potential $\mu $.
  We see that reweighting (black points) and LLR (red points) give compatible results in the region
  where the phase is not too small, but reweighting fails in the strong sign problem
  region.
  Right: natural logarithm of $\langle \e^{i\phi}
    \rangle$ for different values of $\mu$, 
    Statistical errors are smaller than the symbols. The colour code is as
    follows: the plain blue points  (between $\mu=1.0621$
    and $\mu= 1.3721$) have a $\chi^2$ per degree-of-freedom of
    order one, the light blue points between $10$ and
    $50$, and the white points larger than $50$.
    }
\end{figure*}

See~\cite{Langfeld:2016kty} for a review at this conference.
In a nutshell: we assume that the support of $\rho$ is finite and decompose it in small $n_{\rm int}$ intervals
$[s_0,s_1],[s_1,s_2], \ldots [s_{n_{\rm int}-1}, s_{n_{\rm int}}]$.
On each interval, we approximate the Napierian logarithm of the density by a linear function
\be
\ln \rho (s)  = a_i s + b_i \;, \qquad s \in\left[s_{i-1}, s_i\right]\;,
\ee
assuming continuity: 
\bea
\ln \rho (s) &=& a_i (s-s_{i-1}) + \ln \rho (s_{i-1}) \;,
\hspace{3.4cm} 
s \in\left[s_{i-1}, s_i\right]\;, \\
&=& 
a_i (s - s_{i-1}) + \sum_{k=1}^{i-1} a_k(s_k - s_{k-1}) + \ln \rho(s_0)
\;, \qquad
s \in\left[s_{i-1}, s_i\right]\;,\\
\label{eq:logrho}
&=& a_i (s - s_{i-1}) + \sum_{k=1}^{i-1} a_k \ds 
\;,
\hspace{2.2cm}
\qquad \qquad s \in\left[s_{i-1}, s_i\right]\;.
\eea
In the last equation, we assumed that the size of the interval is constant, $s_{i-1} - s_i = \d$,
and that $\ln \rho(s_0)=0$ (note that any non-zero value $\rho (s_0)$ cancels out in the ratio, see eg
Eq.~\ref{eq:phase1}).
The procedure to compute the coefficients $a_i$ is described in detail
in~\cite{Garron:2016noc} and we will not repeat it here.
We reconstruct the density according to  Eq.~(\ref{eq:logrho})
and fit it to an even-degree polynomial
\be
\ln \rho (s) \rightarrow 
\sum _{n}^{{\rm deg}/2} c_j \, s^{2j} \;,
\qquad {\mbox{ fit range } } 0\le s \le s_{\rm fitmax} \;,
\ee 
because the theory is such that $\ln \rho(-s) = \ln \rho (s)$.
Coming back to the phase factor expectation value,
since we are only interested in the real part,
we are left to compute Eq~\ref{eq:phase1}
where we integrate between $0$ and $s_{\rm max}$ (both in the numerator
and the denominator).
The difficulty is course to evaluate the numerator of Eq.~\ref{eq:phase1}.
Besides the numerical challenge, one also has to ensure that result does not depend on $s_{\rm max}$,
$s_{\rm fitmax}$, $n_{\rm int}$, $\ds$, nor on the degree of polymomial $\rm deg$. 
We investigate this at length in~\cite{Garron:2016noc}. Here we show
our final results in Fig.\ref{fig:phase} for the phase factor expectation value (and its logarithm).
We find that the LLR method works very well in the strong sign problem region,
where the reweighing approach fails.

\section{Moment method: Strategy}
\label{sec:moment}

\begin{figure}[t]
  \includegraphics[width=0.5\textwidth]{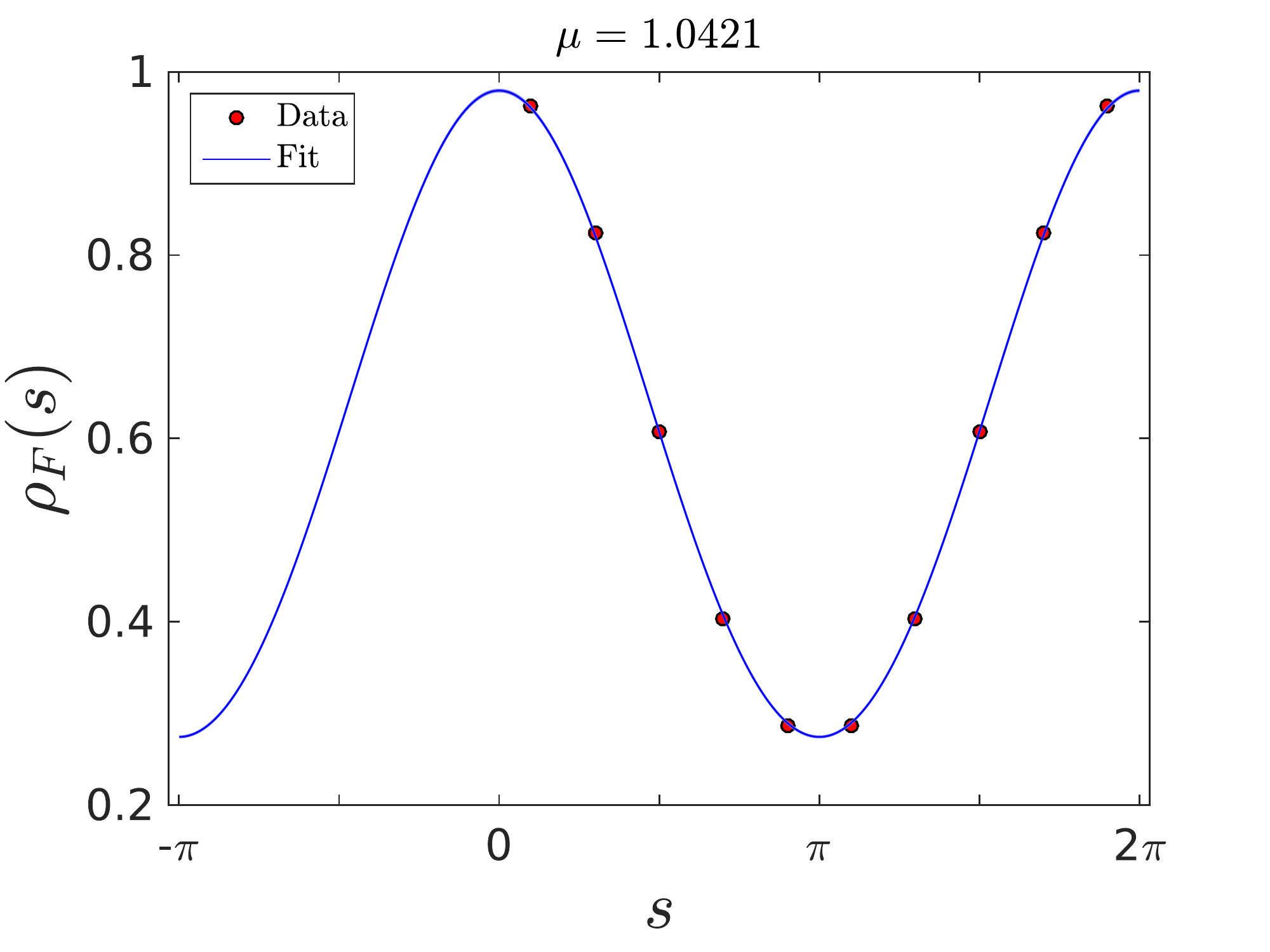}
  \includegraphics[width=0.5\textwidth]{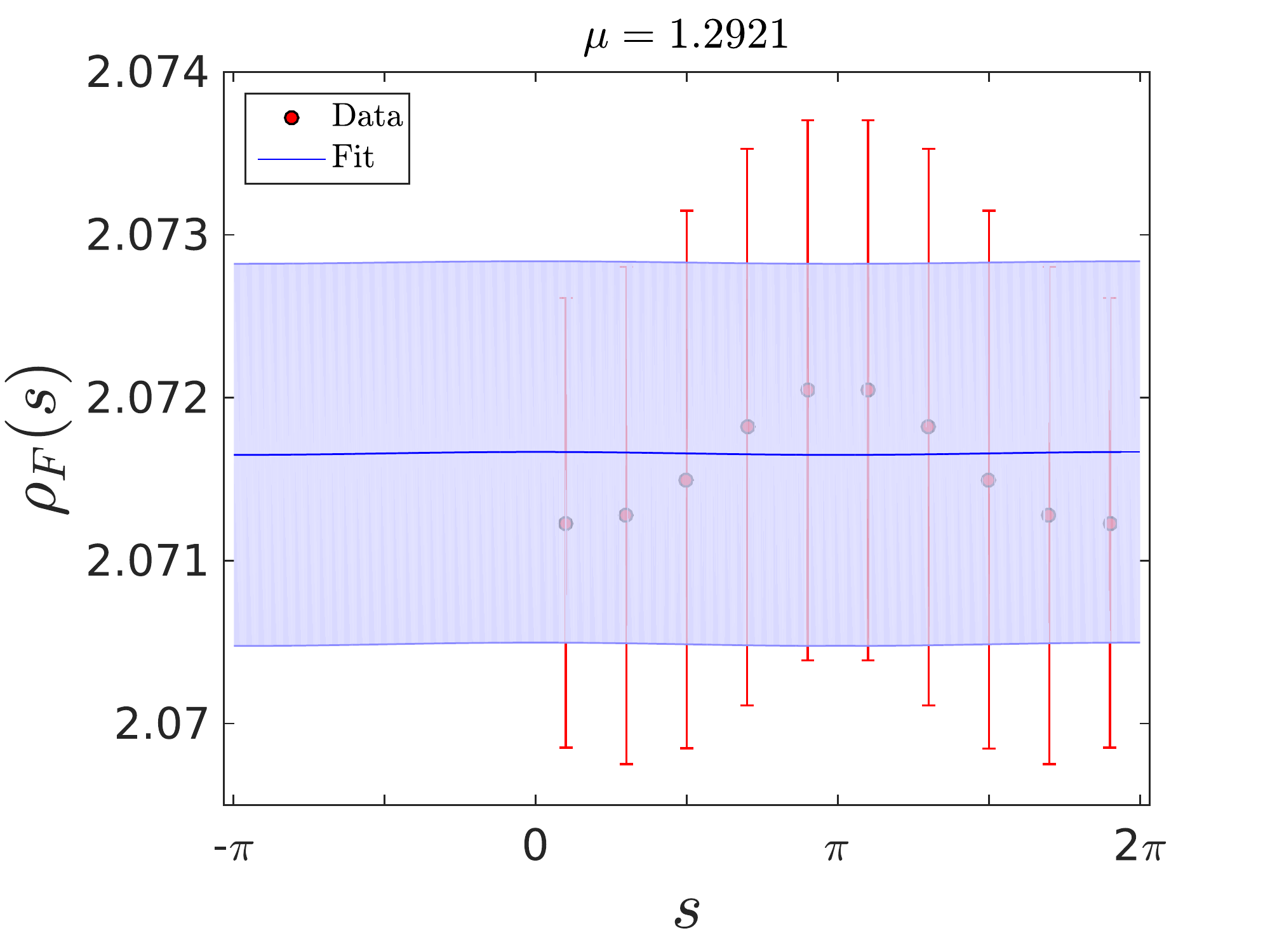}
  \caption{Left: folded density obtained directly from the data
    or from fitting the unfolded density, in the low density-region
    where the sign problem is weak. Right: same but in the strong sign problem regime.
    The blue band corresponds to the $1-\sigma$ region obtained from the fit and
  is almost invisible for $\mu=1.0421$.}
  \label{fig:rhoF1}
\end{figure}

Our moment approach here is similar in strategy to the so-called
cumulant expansion method
(see~\cite{Nakagawa:2011eu,Saito:2012nt,Saito:2013vja}). 
We start with the substitution  $x=s-2k\pi$ 
and splitting the integration domain in intervals of size $2\pi$ gives
\be
\label{eq:phase_rhoF}
\la \e^{i\phi} \ra_{PQ}
= \frac{1}{Z_{PQ}} \sum_{k\in {\mathbb Z}} \int_{-\pi }^{\pi} dx \; \rho(x+2k\pi) \cos(x) 
= \frac{1}{Z_{PQ}} \int_{-\pi }^{\pi} dx \, \rho_F(x) \, \cos(x) \;.
\ee
where we have defined the {\em folded density}: 
$
\rho_F(x) \equiv  \sum_{k\in {\mathbb Z}} \rho(x+2k\pi) \; . 
$
We note that $\rho_F$ can be obtained from the fit of $\rho$ performed in the previous section
but also in an Ansatz-independent way
if the LLR coefficients $a(s)$ are computed at values of $s$
separated by an interval of $\delta_s = 2/(n\pi)$.
We have implemented this technique for HDQCD at several values
of $\mu$ and find that the two methods give compatible results,
within tiny error bars, see Figs.~\ref{fig:rhoF1}.

We propose an expansion in terms of moments which could be seen as a variant of a cumulant
expansion~\cite{Ejiri:2007ga,Nakagawa:2011eu,Saito:2012nt,Saito:2013vja}.
In a first step, we write $\rho_F$ as an even-degree polynomial in $s$:
$
\rho_F(s) = \sum_{k=0}^{N_0} d_k s^{2k}  
$
where chose without a loss of generality $d_0=1$.
We then define the elementary moments defined by
\be
\label{eq:s2n} 
\langle s^{2n} \rangle 
\equiv \frac{1}{Z_{PQ}} \int _{- \pi }^\pi ds \; s^{2n} \; \rho_F(s) 
=
\frac{1}{Z_{PQ}} \, \sum_{k=0}  \frac{2 \pi^{2(n+k)+1}}{2(n+k)+1}d_k \;,
\ee
Plugging $\rho_F(s)  $ in Eq.~\ref{eq:phase_rhoF} gives
\be
\label{eq:expansion1}
\la  \e^{i\phi}\ra  = \frac{1}{N} \sum_{k=1} d_k \, I_{2k} \;,
\quad \mbox{ where }
I_{2k} = \int _{-\pi}^\pi ds \, s^{2k} \,\cos (s) 
= \sum_{l=1}^{k}   (-1)^{k-l+1}  \, \frac{2(2k)!}{(2l-1))!} \, \pi^{2l-1}\;.
\ee

In a second step, we write the phase factor expectation value as a
polynomial in moments:
$
\langle \e^{i\phi} \rangle = \sum_{j=1}^{N_o+1} k_j \langle s^{2i}\rangle   \;.
$
We then impose Eq~\ref{eq:expansion1} to match the latter  equation 
at a given order in $N_o$, for any value of $d_i$.
For example at leading order, $N_o=1$, this gives
\be
-d_1 4\pi = k_1 (\frac{2\pi^3}{3} + d_1 \frac{2\pi^5}{5} ) + k_2 (\frac{2\pi^5}{5} + d_1 \frac{2\pi^7}{7})\;.
\ee
Solving for any $d_1$ gives $k_1 = \frac{105}{2\pi^4}$ and $k_2=\frac{-175}{2\pi^6}$,
therefore
\be
\langle \e^{i\phi} \rangle_{PQ} = -\frac{175}{2\pi^6}
\left(
\langle s^4 \rangle - \frac{ 3 \pi^2 }{5} \, \langle s^2 \rangle 
\right)
\;,
\ee
The generalisation to higher orders is straightforward:
\be
\label{eq:Zexp}
\langle \e^{i\phi} \rangle_{PQ} = \sum_{j=1}^{N_o+1} \alpha_{2(j+1)} M_{2(j+1)}\;,
\ee
where we have introduced the {\em advanced moments} $M_i$.
We give explicitly the first three orders:
\bea
\label{eq:M4}
M_4 &=& \langle s^4 \rangle - \frac{ 3 \pi^2 }{5} \, \langle s^2
\rangle \;, \hbo 
\\
\label{eq:M6}
M_6 &=&
\langle s^6 \rangle \; - \; \frac{ 10 \pi^2 }{9} \, \langle s^4 \rangle \;
+ \; \frac{ 5\pi^4 }{21} \, \langle s^2 \rangle \; , \\
\label{eq:M8}
M_8 &=& \langle s^8 \rangle \; - \; \frac{ 21 \pi^2 }{13} \, \langle s^6 \rangle \;
+ \; \frac{ 105\pi^4 }{143} \, \langle s^4 \rangle \;
- \; \frac{ 35\pi^6 }{429} \, \langle s^2 \rangle \;.
\eea
The $M_i$ depend on the theory under investigation,
as they depend on the elementary moments.
On the other hand, the $\alpha_i$ are constant. Their values for the first orders are :
\bea
\alpha_4 &=& - \frac{175}{2 \pi ^6}
= -0.09101412880 \;,\\
\alpha_6 &=& \frac{4851 \, ( 27 - 2 \pi^2) }{ 8 \pi^{10}} 
=  0.04701392470 \;,\\
\alpha_8 &=& -\frac{57915\,  (3\pi^4 -242 \pi^2 +2145 ) }{16 \pi ^{14} }
= -0.01935715049 \;.
\eea
One should note that each term $\alpha_i M_i$ is positive (see below).

\section{Moment method: Numerical Results for HDQCD}
\begin{figure}[t]
  \begin{center}
    \begin{tabular}{cc}
      \includegraphics[width=0.5\textwidth]{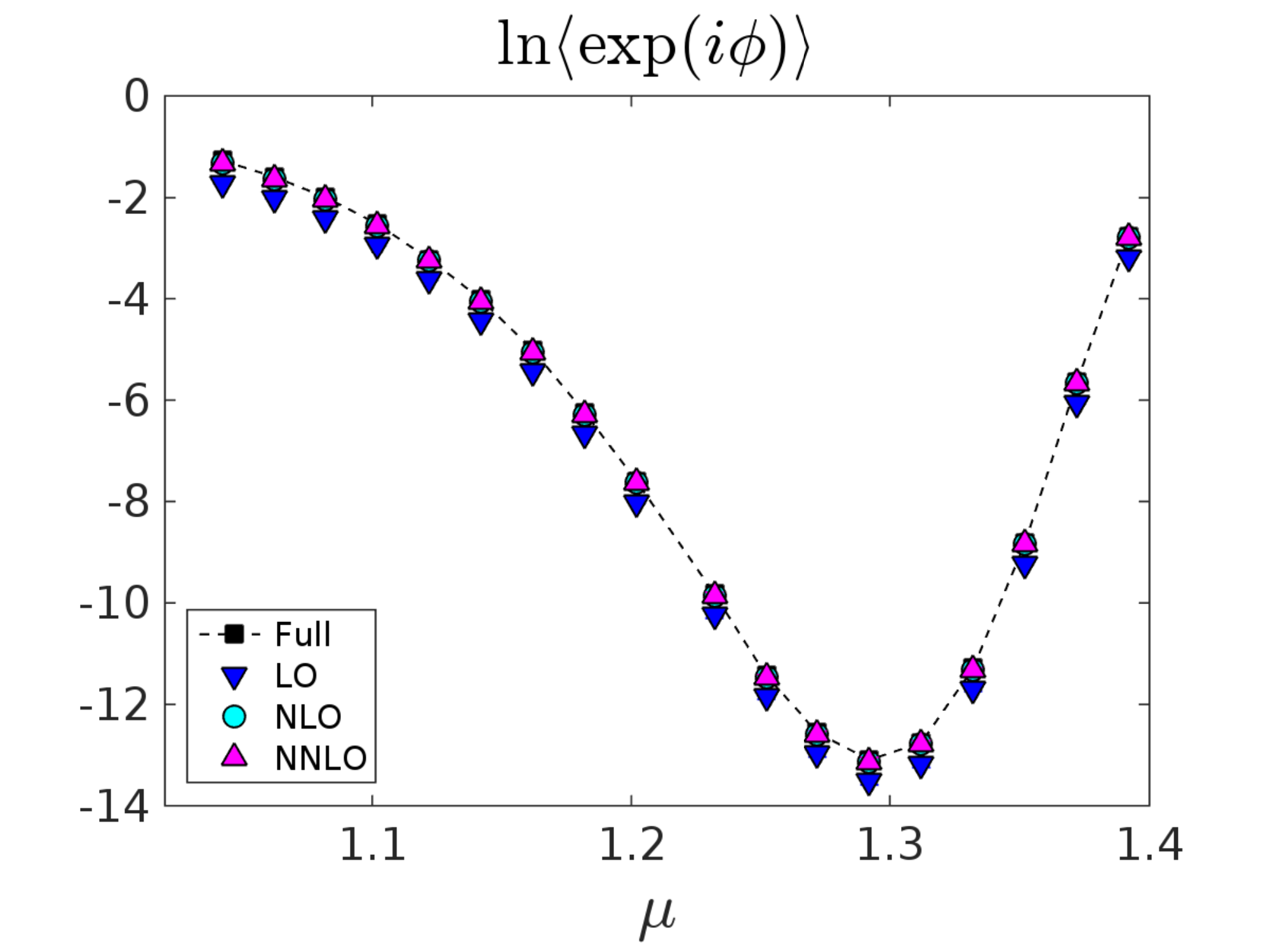} &
      \includegraphics[width=0.5\textwidth]{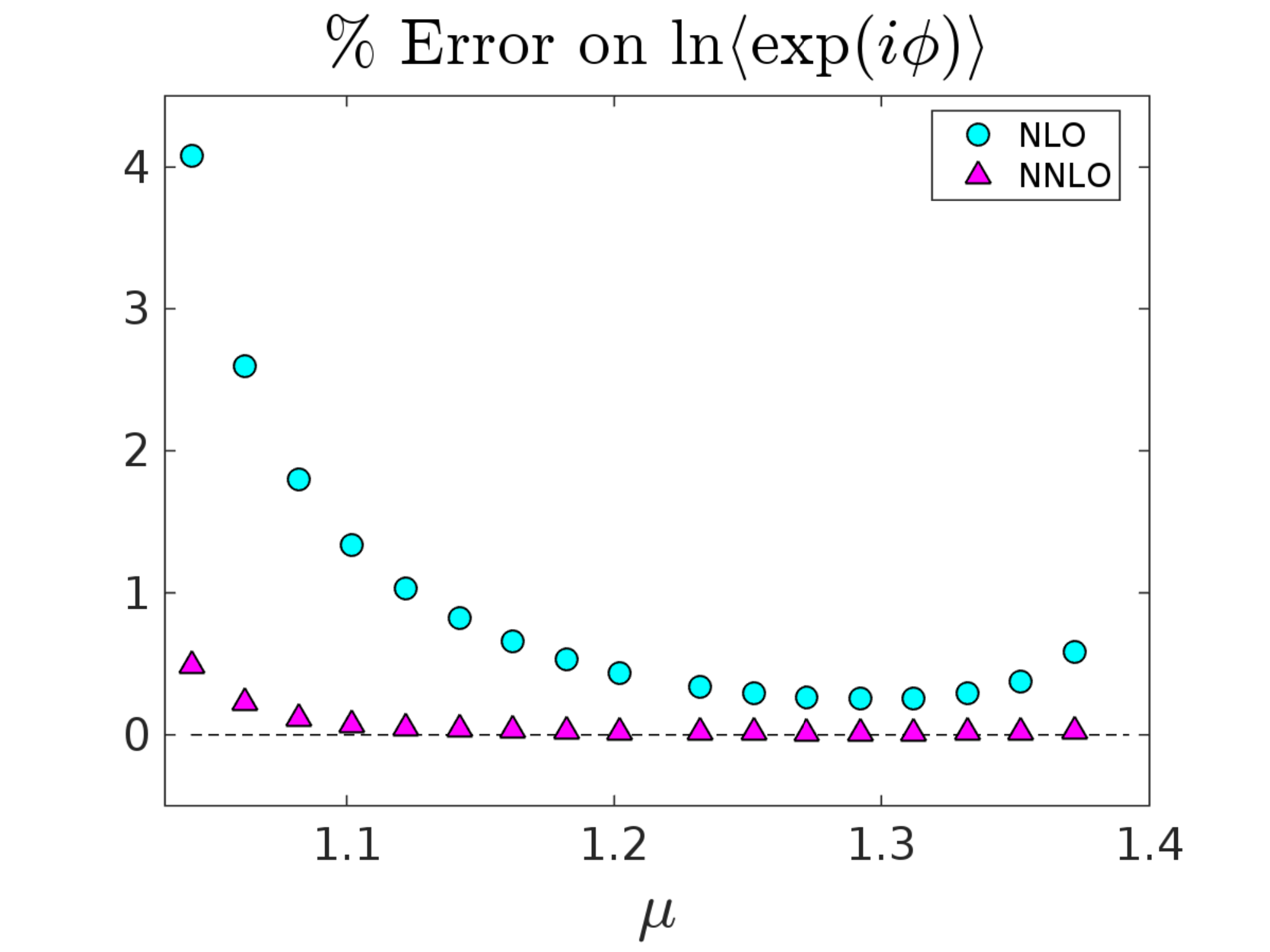}
    \end{tabular}  
  \caption{Left: Comparison of the phase factor expectation value
    computed with the full theory of with the moments method.
    We observe that the moment expansion converges very quickly,
    the NNLO and NNLO lie on top of each other and are indistinguishable
    from the full answer. Statistical error bars are included.
    Right: Relative difference between the full answer and the moment expansion.
    A NNLO, the moment method agrees with the full answer
    at the percent level, and at NNLO order at the sub-percent level or better.}
  \label{fig:phasemoments}
  \end{center}
\end{figure}

In order to test the moment expansion, we apply it to HDQCD,
since we have computed the density using LLR.
For the sake of illustration, we focus on $\mu=1.2921$ where the sign problem is severe.
The folded density is shown in Fig.~\ref{fig:rhoF1}
(we have computed the folded dentisty for $s\in\left[0,2\pi\right]$
rather than $\left[-\pi,\pi\right]$).
The values of the first elementary and advanced moments are reported in Table~\ref{table:moments}.
Thanks to the LLR method, the elementary moments are extracted with a very good statistical precision.
This precision is needed to extract the advanced moments in which important cancellations occur:
\begin{table}[t]
  \scriptsize
  \begin{minipage}{0.55\linewidth}
  \begin{tabular}{cccc}
    \hline \\[-2ex]
    Moment & Central Value & Abs. Err. & Rel. Err. ($\%$) \\[1ex]
    \hline \\[-2ex]
    $\la s^2 \ra$ & $\z\z\z3.\,289\,859\, 4$ & $\z\z  13 \times 10^{-7}$  & $3.9 \times 10^{-5}$\\
    $\la s^4 \ra$ & $\z\z 19.\,481\,750\, 1$ & $\z   100 \times 10^{-7}$  & $5.1 \times 10^{-5}$\\ 
    $\la s^6 \ra$ & $\z  137.\,340\,787\, 5$ & $\z   778 \times 10^{-7}$  & $5.7 \times 10^{-5}$\\ 
    $\la s^8 \ra$ & $   1054.\,276\,996\, 8$ & $    6251 \times 10^{-7}$  & $5.9 \times 10^{-5}$\\
    \hline
  \end{tabular}
  \end{minipage}
  \begin{minipage}{0.45\linewidth}
  \begin{tabular}{cccc}
    \hline \\[-2ex]
    Moment & Central Value & Abs. Err. & Rel. Err. ($\%$) \\[1ex]
    \hline \\[-2ex]
    $M_4$ & $ -1.592\times 10^{-5}$ & $2.35\times 10^{-6}$ & $15\%$ \\
    $M_6$ & $\m1.424\times 10^{-5}$ & $2.11\times 10^{-6}$ & $15\%$ \\
    $M_8$ & $ -3.502\times 10^{-6}$ & $5.18\times 10^{-7}$ & $15\%$ \\
    \hline
  \end{tabular}
  \end{minipage}
  \caption{First elementary (left) and advanced moments (right) with their errors at $\mu=1.2921$.}
  \label{table:moments}
\end{table}
For example, at LO:
\be
\label{eq:M4cancel}
M_4 
= 19.\,481\,750\,4(100) - 19.\,481\,766\,3 (77)  
= -0.\,000\,015\,9(23) \;,
\ee
It is clear from these numbers that not only
the elementary moments have be determined with high precision,
but also their correlations have to be known.
For the phase, Eq.~(\ref{eq:Zexp}) gives
\be
\nn
\la \e^{i\phi} \ra_{PQ}
= 10^{-6} \times \big(
     \underbrace{1.45(21)}_{LO}  
     +  \underbrace{0.67(10)}_{NLO}
     +  \underbrace{0.068(10)}_{NNLO} \,
     \big)
      = 2.186 (323) \times 10^{-6} + {\cal O}(\alpha_{10}M_{10})\;,
\ee
to be compared to the ``real'' answer
$ \la \e^{i\phi} \ra = 2.189(323) \times 10^{-6} $~\cite{Garron:2016noc}.
Since the phase factor is a small number,
it is useful to look at the logarithm of this quantity.
We find 
\bea
\ln \la \e^{i \phi} \ra_{PQ}
&=& -13.043  \pm 0.167 
\hspace{2.6cm} \mbox{(``exact'')} \;,\\
&=& -13.456 \pm 0.167 + {\cal O}(\alpha_{6}  M_6) \qquad (\rm LO) \;,\\
&=& -13.076 \pm 0.167 + {\cal O}(\alpha_{8}  M_8) \qquad (\rm NLO) \;,\\
&=& -13.045 \pm 0.167 + {\cal O}(\alpha_{10} M_10) \quad (\rm NNLO).
\eea
We see that the results converge quickly to the true answer.
It is remarkable that not only the central value but also the variance is very
well approximated by our expansions.
Indeed for this value of $\mu$,
the full (relative) variance of the phase is already given by the first order.
Of course the quality of the approximation depends
on the shape of $\rho_F$ (and therefore on the strength of the sign problem).
As we see in Fig.~\ref{fig:rhoF1}, for $\mu=1.2921$, $\rho_F$ is very well approximated
by a constant.
We now vary the value of $\mu$ in the range $1<\mu<1.4$
and compare the results of the phase factor expectation value obtained
in~\cite{Garron:2016noc} to the method proposed here.
It is interesting to note that even in the weak sign-problem region,
in which the density $\rho_F$ fluctuates between $0$ and $1$,
the NLO and NNLO approximation work remarkably well
(at the per-mille  over the full available range for NNLO).
This is illustrated in Fig.~\ref{fig:phasemoments}.
This method is currently under investigation~\cite{momentpaper}.

\medskip 
{\bf Acknowledgements: } It is pleasure to thank Biagio Lucini,
Roberto Pellegrini and Antonio Rago for discussions. 

\bibliography{mybibfile}
\bibliographystyle{h-elsevier}

\end{document}